\documentstyle[12pt]{article}
\begin{document}
\setlength{\baselineskip}{3.0ex}
\vspace*{2.3cm}

\def\dpi{\bar{B}\to D\pi\ell\bar{\nu}}
\def\dstar{\bar{B}\to D^*\pi\ell\bar{\nu}}
\def\dw{\Gamma_{D^*}}
\def\be{\begin{eqnarray}}
\def\xip{{\Xi'}}
\def\en{\end{eqnarray}}
\def\es{\varepsilon}
\def\ri{\to}
\def\pr{{\em Phys. Rev.}~}
\def\prl{{\em Phys. Rev. Lett.}~}
\def\np{{\em Nucl. Phys.}~}
\def\pl{{\em Phys. Lett.}~}
\def\zp{{\em Z. Phys.}~}

{\obeylines
\hfill ITP-SB-93-41}
\vskip 0.3 cm
\begin{center}
{\large\bf Predictions for Semileptonic Decays $\bar{B}\to D(D^*)\pi\ell\bar{
\nu}$ and Weak Radiative Decays of Bottom Hadrons
\footnote{Talk presented at the {\it 5th International Symposium on Heavy
Flavour Physics}, Montreal, July 6-10, 1993.}
}\\
\vskip 1.1cm
{\large Hai-Yang Cheng} \\
\vspace*{1.5ex}
{\large\it Institute of Physics, Academia Sinica} \\
{\large\it Taipei, Taiwan 11529, R.O.C.}\\
\vskip 0.2 cm
\centerline{and}
\vskip 0.3 cm
{\large\it Institute for Theoretical Physics} \\
{\large\it State University of New York} \\
{\large\it Stony Brook, NY 11794} \\
\vspace*{1.5ex}
%
%
%
%
\vspace*{6.0ex}
%
%
%
%
\end{center}
\vspace*{6.0ex}
\begin{center}
{\bf Abstract}
\end{center}

\noindent The semileptonic decays $\bar{B}\to D(D^*)\pi\ell\bar{\nu}$ are
analyzed by heavy quark and chiral symmetries. The branching ratio is of order
of a few percent for $\dpi$, but it is much smaller for $\dstar$, of order
$10^{-4}-10^{-5}$. The decay mode $B^-\to D^+\pi^-e^-\bar{\nu}_e$ provides
a nice place for extracting the $D^*D\pi$ coupling constant from the
semileptonic decays of a $\bar{B}$ meson. Two types of weak radiative decays
of $\bar{B}$ mesons and bottom baryons are studied. The decay $\Lambda_b^0\to
\Lambda^0\gamma$ receives the electromagnetic penguin $b\to s\gamma$
contribution and has a branching ratio
of order $2\times 10^{-5}$. The radiative decays $\bar{B}\to D^*\gamma,~
\Lambda_b^0\to\Sigma_c^0\gamma,~\Xi_b^0\to
\Xi_c^0\gamma$ and $\Xi_b^0\to{\Xi'}_c^0\gamma$ occur through $W$ exchange
accompanied with a photon emission and their branching ratios are
estimated to be much smaller.
We conclude that the weak radiative decays of bottom hadrons are dominated
by the short-distance electromagnetic penguin mechanism.

\pagebreak
\section{Semileptonic Decays $\bar{B}\to D(D^*)\pi\ell\bar{\nu}$}

An ideal theoretical framework for studying the interactions of heavy hadrons
with soft Goldstone bosons is provided by the effective Lagrangians which
incorporate both heavy quark and chiral symmetries [1,2]. When supplemented
by the nonrelativistic quark model, the formalism determines completely the
low-energy
dynamics of heavy hadrons. It turns out that the four-body semileptonic
decays of a $\bar{B}$ meson such as $\dpi$ and $\dstar$ furnish a best place
for testing the synthesis of spin-flavor symmetry of heavy quarks and chiral
symmetry of light quarks. The semileptonic decays with a soft pion are
completely fixed by the Isgur-Wise function measured in $\bar{B}\to D^*\ell
\bar{\nu}$ and the coupling constant that describes the strong decay $D^*\to
D\pi$ [3].

Let us first consider the decay $\dpi$. It proceeds through $D^*$ and $\bar{B}
^*$
pole diagrams. Since the intermediate $D^*$ can be on its mass shell, it
is necessary to use the full propagator for $D^*$ to incorporate its finite
width $\dw$. Results for the decay rates are separated into two
categories: resonant and nonresonant. The resonant part is defined as those
events with the invariant mass of $D\pi$ satisfying $|m_{D\pi}-m_{D^*}|<3\dw$.
All others are nonresonant. We have found that the contributions from the $D^*
$ pole dominate both the resonant and nonresonant decays. Since the validity
of chiral symmetry demands the emitted pions be soft, we have to impose
cutoffs on the pion momenta in our calculation. We simply cut off the pion's
3-momentum at 100 MeV or 200 MeV in the appropriate frame of reference.

The integrated decay rates of $\dpi$ in the resonant and nonresonant
regions are sensitive to the total decay width of $D^*$. By fixing the
$D^*D\pi$ coupling constant and treating $\dw$ as a free parameter, we find a
linear relationship between the integrated rate and $1/\dw$. More precisely,
\be
\Gamma(\dpi)=\,f^2\left({A\over \dw}+C\right),
\en
where $f$ is the $D^*D\pi$ coupling constant (it is sometimes denoted by $g$,
which is related to $f$ by $f=-2g$), and the constant terms $A$ and $C$ are
independent
of $\dw$. The constant $A$ arises from the $D^*$ pole, while $C$ comes from
the nonleading contributions of the $D^*$-pole diagram, the $\bar{B}^*$-pole
contributions, and the interference terms between the $\bar{B}^*$ and $D^*$
pole diagrams. The constants $A$ and $C$ are generally very different for
resonant and nonresonant contributions.

Throughout our calculations, we have used the meson Isgur-Wise function
proposed in Ref.[4]
\be
\xi(y)=\,1-\rho^2(y-1)+c(y-1)^2,
\en
where the parameters $\rho$ and $c$ are determined by fitting Eq.(2) to the
measured $\bar{B}\to D^*\ell\bar{\nu}$ spectra to be $\rho=1.08\pm 0.10$ and
$c=0.62\pm 0.15$ [4]. As for the $D^*D\pi$ coupling, we use $f=-1.50$ (or
$g=0.75$) inferred from the chiral quark model, which in turn gives rise to the
correct value of $g_A^{\rm nucleon}=1.25$ [2]. We have applied chiral symmetry
and the nonrelativistic quark model to calculate the strong and radiative
decays of $D^*$ [5]. Our predictions for the branching ratios of $D^*\to D\pi$
and $D^*\to D\gamma$ are in excellent agreement with the most recent CLEO II
experiment [6]. Our predicted total widths are [5]
\be
\Gamma(D^{*+})=141\,{\rm keV},~~~~\Gamma(D^{*0})=102\,{\rm keV}.
\en

   We have computed the single particle spectra for $D$, the electron and the
pion. The reader is referred to Ref.[3] for detail. In the following we
discuss the essential physics. (i) Although the shape of the single-particle
spectra in the resonant and nonresonant regions looks similiar (see Figs.4-6
of Ref.[3]), the rates in the resonant region are larger than those in the
nonresonant region by a factor of 7. The nonresonant decay is insensitive
to the pion momentum cutoff. For example, the decay rate increases by only
15$\%$ when the cutoff increases from 100 MeV to 200 MeV. (ii) The linear
relation between $\Gamma(\dpi)$ and $1/\dw$ for both resonant and nonresonant
decays, as expected theoretically in Eq.(1), is borne out by our numerical
work (see Figs.7-10 of Ref.[3]). Therefore, equipped with our theoretical
results for $\Gamma(D^{*\pm})$ and $\Gamma(D^{*0})$, we are able to predict
the decay rates of $\bar{B}\to(D\pi)_{\rm res}\ell\bar{\nu}$ and
$\bar{B}\to (D\pi)_{\rm nonres}\ell\bar{\nu}$. The numerical results show that
the straight lines for the resonant contributions for both charged and
neutral $\bar{B}$ mesons pass through the origin. This implies that $C_{\rm
res}\approx 0$, as expected owing to a very small phase space contributing
to the constant $C$. We will discuss its implication shortly. Contrary to
the resonant part, we find $C\neq 0$ in the nonresonant region.
(iii) The decay mode $B^-\to D^+\pi^-e^-\bar{\nu}_e$ deserves special
attention. Because $m_{D^{*0}}<m_{D^+}+m_{\pi^-}$, its resonant decay rate
is completely negligible. Moreover, the rate for $B^-\to (D^+\pi^-)_{\rm
nonres}e^-\bar{\nu}_e$ is nearly independent of $\dw$. (iv) If we identify
the $D\pi$'s in the resonant region with the $D^*$, we can even predict the
decay rate of $\bar{B}\to D^*\ell\bar{\nu}$ provided that the branching ratio
of $D^*\to D\pi$ is known, as in our case. Using the predicted branching
ratio of
$66.7\%$ for $D^{*0}\to D^0\pi^0$, we find $B(B^-\to D^{*0}e^-\bar{\nu}_e)=
5.23\%$, in good agreement with experiment. This represents a trumph for heavy
quark symmetry as it is independent of the chiral symmetry of light quarks.
(v) Numerically, the branching ratio is of order $(2-3)\%$ for $\bar{B}\to(D
\pi)_{\rm res}\ell\bar{\nu}$,
while it ranges from $0.4\%$ to $0.7\%$ for
nonresonant decays, depending on the pion momentum cutoff.

   It has been suggested that the decay rate for $\dpi$ can be used to fix the
$D^*D\pi$ coupling constant [7,8]. Authors of [7] proposed to constrain the
coupling $f$ from the resonant decays $\bar{B}\to(D\pi)_{\rm res}\ell\bar{
\nu}$. However, as we pointed out before, the $\bar{B}^*$ pole contribution
is negligible due to $C_{\rm res}\approx 0$. Since the charged $D^{*+}$ decays
almost exclusively to $D\pi$, $\Gamma(D^{*+})$ is governed by $f^2$. It is
easily seen from Eq.(1) that the decay rate for $\bar{B}\to D^{*+}\ell\bar{
\nu}\to(D\pi)^+\ell\bar{\nu}$ is essentially independent of $f^2$. The neutral
$D^{*0}$ has a substantial radiative decay contribution [5,6], so $\Gamma(D^{
*0})$
is not simply related to $f^2$. Therefore, we conclude that it is impossible
to determine the $D^*D\pi$ coupling constant from $\bar{B}\to(D\pi)_{\rm res}
\ell\bar{\nu}$ without having other information on $D^*$ decay. Authors of [8]
considered the kinematic region away from the $D^*$ pole in $\dpi$ in order to
fix $f$.
As noted in passing, the rate of $B^-\to D^+\pi^-e^-\bar{\nu}_e$ is nearly
independent of $\Gamma(D^{*0})$ and hence is proportional to $f^2$.
 Consequently, the best place for extracting the $D^*D\pi$ coupling
constant from semileptonic decays of a $\bar{B}$ meson lies in the decay $B^-
\to D^+\pi^-e^-\bar{\nu}_e$.

   We next turn to the decay $\dstar$ (for a previous study, see [9]).
Depending on the pion momentum cutoff scheme, the overall branching ratios are
of order $10^{-4}-10^{-5}$, which are smaller than $\dpi$
by two to three orders of magnitude. This is ascribed to the fact that none
of the $\bar{B}^*,~D^*$ and $D$ intermediate states in the pole diagrams can
be on or close to its
mass shell. The polarization of $D^*$ is a new feature of this decay. We have
studied the single particle spectra for each polarization of $D^*$ in the
$\bar{B}$-meson rest frame. In all cases, contributions from the left-handed
and longitudinal polarizations dominate that from the right-handed
polarization. This can be understood as a result of the $V-A$ coupling of the
quarks to the $W^\pm$ bosons. The charmed quark produced by the $\bar{B}$
decay is predominately left-handed. Therefore, a study of the single particle
spectra for different $D^*$ polarizations can be used to exploit the nature of
weak interaction dynamics.

\section{Weak Radiative Decays of Bottom Hadrons}
   Motivated by the recent observation of the decay $\bar{B}\to K^*\gamma$ by
CLEO [10], we have systematically analyzed the two-body weak radiative decays
of $\bar{B}$ mesons and bottom baryons using heavy quark symmetry and the
nonrelativistic quark model [11]. The measured branching ratio of $(4.5\pm
1.5\pm 0.9)\times
10^{-5}$ for $\bar{B}\to K^*\gamma$ confirms the standard-model expectation
that this decay mode is dominated by the short-distance electromagnetic
penguin transition $b\to s\gamma$. Other
two-body radiative decays of bottom hadrons proceeding through the $b\to s
\gamma$ process are
\be
 \bar{B}_s\ri\phi\gamma,~~~
 \Lambda_b^0 \ri \Sigma^0\gamma,~\Lambda^0\gamma,~~~\Xi^0_b\ri\Xi^0\gamma,
{}~~~\Xi^-_b\ri\Xi^-\gamma,~~~\Omega_b^-\to\Omega^-\gamma.
\en
Another type of radiative decays occurs through $W$ exchange accompanied with
a photon emission. Examples are
\be
 \bar{B} \ri D^*\gamma,~~~\bar{B}_s\ri D^*\gamma,~~~
\Lambda_b^0  \ri \Sigma^0_c\gamma,~~~\Xi^0_b\ri\Xi^0_c\gamma,~{\Xi'}^0_c\gamma.
\en

   We first concentrate on the radiative decay $\bar{B}\to D^*\gamma$. Our
goal is to see if the long-distance effect due to the tree-level $W$-exchange
with a photon emission is competitive with the short-distance one-loop $b
\to s\gamma$ mechanism. The long-distance contribution is usually evaluated
under the pole approximation, namely it is saturated by one-particle
intermediate states. For $\bar{B}\to D^{*0}\gamma$ decay, there are three
pole diagrams with $D^0,~D^{*0}$ and $\bar{B}^{*0}$ intermediate states.
Unfortunately, one cannot apply heavy quark symmetry or the quark model
directly to the electromagnetic vertices in the pole diagrams. This is because
the intermediate states are very far away from their mass shell.
For example, the four-momentum squared of the $D$ pole
is $m^2_B$. This means that the residual momentum of the
$D$ meson defined by $P_\mu=m_Dv_\mu+k_\mu$ must be of order $m_B$, so
the approximation $k/m_D<<1$ required by the heavy quark effective theory is
no longer valid. Nevertheless, we can factorize the off-shell photon coupling
as, for instance,
\be
g_{_{DD^*\gamma}}(q^2=m^2_B)=\,g_{_D}(q^2=m^2_B)g_{_{DD^*\gamma}},
\en
where $g_{_{DD^*\gamma}}$ is an on-shell photon coupling constant, and $g_{_D}$
is a form factor accounting for off-shell effects. The form factor is
normalized to unity when particles are on shell, $g_{_D}(q^2=m^2_D)=1$. As a
consequence, we can still apply heavy quark symmetry and the quark model
to the on-shell photon coupling constants, as elaborated on in Ref.[5].
Although the
form factors such as $g_{_D}(q^2)$ are basically unknown, they are expected
to become smaller as the intermediate pole state is more away from its mass
shell owing to less overlap of initial and final hadron wave functions. We
then employ the QCD-corrected effective weak Hamiltonian to evaluate the
relevant weak matrix elements.

   Referring the calculational detail to Ref.[11], we simply write down the
final result
\be
B(\bar{B}^0\to D^{*0}\gamma)=2\times 10^{-5}\left\{\left[g_{_D}(m^2_B)
-0.16g_{_{B^*}}(m^2_D)\right]^2+0.32g^2_{_{D^*}}(m^2_B)\right\}.
\en
It is evident that the upper bound for the branching ratio of $\bar{B}^0\to
D^{*0}\gamma$ is $2\times 10^{-5}$. If the form factor proposed in [12]
$g_{_D}(q^2)=(m^2_{D'}-m^2_D)/(m^2_{D'}-q^2)$ ($D'$ being the first radial
excitation of the $D$ meson) is used, the branching ratio will be only
of order $10^{-7}$. The suppression of $\bar{B}\to D^*\gamma$ relative to
$\bar{B}\to K^*\gamma$ is mainly attributed to the smallness of the decay
constants
$f_D$ and $f_B$ occurred in weak transitions. Two remarks are in order. (1)
A sizeable long-distance contribution to $\bar{B}\to K^*\gamma$ via the weak
transition $\bar{B}\to K^*\psi$ followed by an electromagnetic conversion
$\psi-\gamma$ has been advocated in the past [13]. This mechanism, if works,
could be the dominant effect for $\bar{B}\to D^*\gamma$ decay. Unfortunately,
there is a gauge invariance problem with the parity-violating amplitude.
Gauge invariance requires that the two axial form factors $A_1$ and $A_2$
appearing in $\bar{B}-V$ transition ($V=\rho,\,\omega$ for $\bar{B}\to D^*
\gamma$ and $\psi$ for $\bar{B}\to K^*\gamma$) be the same. However, it
is easy to check that $A_1\neq A_2$ in the limit of heavy quark symmetry.
We thus believe that it is not pertinent to apply the vector-meson-dominance
model in the present form to the study of weak radiative decays.
(2) A previous quark-model
calculation in [14] gives $B(\bar{B}^0\to D^{*0}\gamma)\sim 10^{-6}$.
\footnote{This number is obtained after a correction with the updated values
of $f_B,~f_{D^*},~V_{cb}$ and a replacement of $(2c_+-c_-)/3$ by $(c_+-c_-)
/2$. I thank R.R. Mendel for communication on this.}

   Since the weak radiative decay of $\bar{B}$ mesons is dominated by the
electromagnetic penguin diagram, it is natural to expect that the same
mechanism dominates in bottom baryon decays. We consider the short-distance
effect in the decays $\Lambda_b^0\to\Sigma^0\gamma$ and $\Lambda_b^0
\to\Lambda^0\gamma$ by first treating the $s$ quark as a heavy quark and then
taking into account the $1/m_s$ and QCD corrections. Despite that the effective
mass of the $s$ quark is only around 500 MeV, it is not small compared to
the QCD scale and we thus expect to see some vestiges of heavy quark symmetry.
In the heavy $s$ quark
limit, the hyperon $\Lambda$ behaves as an antitriplet heavy baryon
$B_{\bar{3}}$, while $\Sigma^0$ as a sextet baryon $B_6$. It turns out that
the weak $B_{\bar{3}}-B_6$ transition is prohibited in the heavy quark limit,
as we noted before [15].
So, our first prediction is $\Gamma(\Lambda_b^0\to\Sigma^0\gamma)<<\Gamma(
\Lambda_b^0\to\Lambda^0\gamma)$.
A detailed calculation yields the amplitude $A(\Lambda_b^0\to\Lambda^0\gamma)=
\,i\bar{u}_\Lambda(a+b\gamma_5)\sigma_{\mu\nu}\es^\mu q^\nu u_{\Lambda_b}$
with
\be
a=b={G_F\over \sqrt{2}}\,{e\over 8\pi^2}F_2m_bV_{tb}V^*_{ts}
\left(1-{\bar{\Lambda}\over 2m_s}\,{1-v\cdot v'\over 1+v\cdot v'}\right)
C(\mu)\zeta(v\cdot v',\mu),
\en
where $\bar{\Lambda}=\,m_{_{\Lambda_b}}-m_b=\,m_{_{\Lambda_c}}-m_c=\,m_{
\Lambda}-m_s\approx 700$ MeV, $F_2=0.73$ for $m_t=150$ GeV,
and $C(\mu)$ is a QCD-correction factor. We find
that the $1/m_s$
correction to the amplitude is about $30\%$. Using the two recent models [16]
for the Isgur-Wise function $\zeta(v\cdot v')$ in $\Lambda_b\to\Lambda_c$
transition, we arrive at (for $\tau(\Lambda_b)=1.2\times 10^{-12}s$)
\be
B(\Lambda^0_b\to\Lambda^0\gamma)=\,1.34\times 10^{-3}|\zeta(v\cdot v'=2.55)|^2
=\,(1.3-2.3)\times 10^{-5},
\en
which is of the same order of magnitude as $\bar{B}\to K^*\gamma$. In view of
theoretical uncertainties involved, the prediction (9) might be regarded as
a benchmarked value.

   We then proceed to the decays $\Lambda_b^0\to\Sigma_c^0\gamma,~\Xi_b^0\to
\Xi_c^0\gamma$ and $\Xi_b^0\to{\Xi'}_c^0\gamma$ (${\Xi'}_Q$ denotes a sextet
heavy baryon) by considering their long-distance pole contributions. Just as
the case of $\bar{B}\to D^*\gamma$, we apply the nonrelativistic quark model
and heavy quark symmetry to the electromagnetic vertices by first treating
the intermediate state as being on its mass shell and then using form factors
to account for off-shell effects. Methods of evaluating the baryon-baryon
matrix elements are elaborated on in [15]. Many results obtained there are
still applicable in the present study. For example, to the leading order
in heavy quark symmetry, $B_{\bar{3}}-B_6,~B_6-B_6^*$ and $B_{\bar{3}}-B_6^*$
($B_6^*$ being a spin ${3\over 2}$ sextet baryon) are forbidden. Using the MIT
bag model to evaluate the weak matrix elements, we find
\be
B(\Xi^0_b\to \xip^0_c\gamma)=7\times 10^{-8}|g_{_{\xip_b}}(m^2_{\xip_c})|^2.
\en
This branching ratio with its upper bound being $7\times 10^{-8}$ is
unobservably small due to the smallness of the weak $\Xi'_b-\Xi_c'$ transition.
The same conclusion applies to the radiative decays $\Xi_b^0\to\Xi_c^0\gamma$
and $\Lambda_b^0\to\Sigma_c^0\gamma$.

We conclude that the weak radiative decays of bottom hadrons are dominated by
the short-distance electromagnetic penguin mechanism. This phenomenon is quite
unique to the bottom hadrons which contain a heavy $b$ quark; such a magic
short-distance enhancement due to a large top quark mass and large QCD
corrections does not occur in the systems of charmed and strange hadrons.
\vskip 0.5 cm
Acknowledgments: I wish to thank T. M. Yan, C. Y. Cheung, W. Dimm, G. L. Lin,
Y. C. Lin,
and H. L. Yu for a very enjoyable collaboration. I am also grateful to R. R.
Mendel for bringing [14] to my attention, and to the organizer D. B. MacFarlane
for a fruitful and well organized symposium.

\vskip 0.7 cm
\noindent {\large\bf References}
\vskip 0.6 cm

\begin{enumerate}

\item M.B. Wise, {\em Phys. Rev.} {\bf D45}, R2188 (1992);
G. Burdman and J. Donoghue, {\em Phys. Lett.} {\bf B280}, 287 (1992).

\item T.M. Yan, H.Y. Cheng, C.Y. Cheung, G.L. Lin, Y.C. Lin, and H.L. Yu,
{\em Phys. Rev.} {\bf D46}, 1148 (1992).  See also T.M. Yan, {\em Chin. J.
Phys.} (Taipei) {\bf 30}, 509 (1992).

\item H.Y. Cheng, C.Y. Cheung, W. Dimm, G.L. Lin, Y.C. Lin, T.M. Yan, and H.L.
 Yu, CLNS 93/1204, to appear in Phys. Rev. D.

\item G. Burdman, {\em Phys. Lett.} {\bf B284}, 133 (1992).

\item H.Y. Cheng, C.Y. Cheung, G.L. Lin, Y.C. Lin, T.M. Yan, and H.L. Yu,
{\em Phys. Rev.} {\bf D47}, 1030 (1993).

\item The CLEO Collaboration, F. Butler {\it et al.}, {\sl Phys. Rev. Lett.}
{\bf 69}, 2041 (1992).

\item G. Kramer and W.F. Palmer, \pl {\bf B298}, 437 (1993).

\item Clarence L.Y. Lee, Ming Lu, and Mark B. Wise, {\em Phys. Rev.} {\bf
D46}, 5040 (1992).

\item Clarence L.Y. Lee, CALT-68-1841 (1992).

\item The CLEO Collaboration, R. Ammar {\it et al.,} CLNS 93/1212, CLEO 93-06
(1993).

\item H.Y. Cheng {\it et al.}, IP-ASTP-21-93 (in preparation) (1993).

\item P. Colangelo, G. Nardulli, N. Paver, and Riazuddin, \zp {\bf C45}, 575
(1990).

\item E. Golowich and S. Pakvasa, \pl {\bf B205}, 393 (1988);
N.G. Deshpande, J. Trampetic, and K. Panose, \pl {\bf B214}, 467 (1988);
M.R. Ahmady, D. Liu, and Z. Tao, IC/93/26 (1993).

\item R.R. Mendel and P. Sitarski, \pr {\bf D36}, 953 (1987).

\item H.Y. Cheng, C.Y. Cheung, G.L. Lin, Y.C. Lin, T.M. Yan, and H.L. Yu,
{\sl Phys. Rev.} {\bf D46}, 5060 (1992).

\item E. Jenkins, A. Manohar, and M.B. Wise, \np {\bf B396}, 38 (1993);
M. Sadzikowski and K. Zalewski, preprint (1993).

\end{enumerate}

\end{document}